\documentclass[preprint]{aastex}\usepackage{amssymb,epsfig}
\newcommand{\beq}{\begin{equation}}
\newcommand{\eeq}{\end{equation}}
\newcommand{\bea}{\begin{eqnarray}}
\newcommand{\eea}{\end{eqnarray}}
\newcommand{\rem}[1]{ }
\bibpunct[,]{(}{)}{;}{a}{}{,}

\begin{document}

\title{Radiation of electrons in Weibel-generated fields: a general case}

\author{Mikhail V. Medvedev}

\affil{Department of Physics and Astronomy, 
University of Kansas, KS 66045}

\begin{abstract}
Weibel instability turns out to be the a ubiquitous phenomenon in High-Energy Density environments, ranging from astrophysical sources, e.g., gamma-ray bursts, to laboratory experiments involving laser-produced plasmas. Relativistic particles (electrons) radiate in the Weibel-produced magnetic fields in the Jitter regime. Conventionally, in this regime, the particle deflections are considered to be smaller than the relativistic beaming angle of 1/$\gamma$ ($\gamma$ being the Lorentz factor of an emitting particle) and the particle distribution is assumed to be isotropic. This is a relatively idealized situation as far as lab experiments are concerned. We relax the assumption of the isotropy of radiating particle distribution and present the extension of the jitter theory amenable for comparisons with experimental data.
\end{abstract}

\maketitle

\section{Introduction}

In general, high Mach number shocks, e.g., relativistic shocks of gamma-ray bursts, must be highly turbulent. It has been shown that the Weibel instability \cite{Weibel59,Fried59}  is responsible for the GRB shock formation \citep{ML99} on the microscopic level. This instability is driven by the anisotropy of the particle distribution function (PDF) associated with a large number of particles reflected from the shock into the upstream region. This theoretical prediction has recently been confirmed in a number of state-of-the-art numerical sumulations \citep{Silva+03,Fred+04,Nish+04,Spitk08}. The Weibel  instability is also observed in laser-plasma experiments. In particular, it is the goal of the Hercules experiment at the university of Michigan (\citealp{GRB+Hercules08}; for technical details, see \citealp{Hercules08}) to create and diagnose the Weibel instability and turbulence in the laboratory high-energy density plasmas, as a part of the Laboratory Astrophysics and High-Energy Density Physics programs.

The state of the Weibel turbulence corresponds to the self-organized nonlinear regime of the Weibel instability, which is characterized by reorganization of the currents and magnetic fields via strong interaction and hierarchical merger process of current filaments. The Weibel-generated magnetic fields are very small-scale, of order several plasma skin depths, $c/\omega_p$, which is much smaller than the typical Larmor radius of particles in such fields. Thus, if relativistic electrons are present in such a plasma, they will produce emission that is different from the standard synchrotron radiation. Such radiation, referred to as the ``jitter radiation''  has distinct spectral properties \citep{M00}, has been predicted to be emitted from the Weibel turbulence. It has also been predicted that jitter radiation can explain various observational data from GRBs. At last but not least, jitter radiation has been suggested as an interesting diagnostic of the Weibel turbulence in laser plasma experiments. Here we elaborate more on jitter radiation theory in application to the experiments.

\section {Weibel turbulence}

The instability under consideration was first predicted by \cite{Weibel59} for a non-relativistic plasma with an anisotropic distribution function, and the physical interpretation was provided later by \cite{Fried59}, who considered the extreme case of anisotropy --- two counter-streaming particle (plasma) beams. In essence, the two electron-proton plasma streams experience deflections in seed magnetic fields due to the Lorentz force, $ e({\bf v \times B})/c$, so that protons (and electrons) moving in opposite directions concentrate in spatially separated current filaments. The magnetic field of these filaments appears to increase the initial magnetic field fluctuation. The growth rate and the wavenumber of
the fastest growing mode (which, in fact, sets the spatial correlation
scale of the produced field) are of order the plasma frequency, $\omega_p$, and the plasma skin depth, $c/\omega_p$, respectively. The current in the filaments and the associated magnetic fields increase until the energy density in the fields reaches about $\epsilon_B\sim10\%$ of the kinetic energy density of the streaming particles, which is enough to rapidly isotropize the particle distribution; hence the instability quenches.

At longer times, the plasma with the Weibel-generated currents  and fields enters the turbulent state, referred to as the "Weibel turbulence", in which current filaments begin to interact with each other, forcing like currents to approach each other and merge. The filament coalescence is a hierarchical and self-similar process \citep{M+05}.  For filaments with the initial separation $\sim D_0$  the magnetic field correlation length in the non-relativistic and relativistic filaments regimes are 
\beq
\lambda_B(t)=D_0 2^{{t}/({2\tau_{0,NR}})}, \qquad
\lambda_B(t)\simeq  ct. \label{lambda}
\eeq
Here, the typical non-relativistic time-scale  is determined
by Eq. (\ref{lambda}). The coalescence time may be written as
\beq
\tau_{0,NR}\sim
=\frac{(c/v)}{\sqrt{\epsilon_{B}}\,\omega_p}
\sim 10^4\,\omega_p^{-1},
\label{plas-tau0NR}
\eeq
Here $v$ is the beam velocity and we assumed the typical value: $\epsilon_{B}\sim10^{-1}$. Numerical PIC simulations showed \citep{M+05} that both a non-power-law non-relativistic regime and a power-law regime are clearly present in the dynamics. The power-law fits yield $\lambda_B(t)\propto t^\alpha$ with $\alpha\approx0.8$. It should be noted that the field scale growth is somewhat analogous to the inverse cascade in two-dimensional magnetohydrodynamic (MHD) turbulence, with the crucial difference that the former is an entirely {\em kinetic} process since at such small scales
$\sim c/\omega_p$ the MHD approximation is completely inapplicable.

\section{Jitter radiation}

The jitter regime realizes when the deflection angle, $\alpha$, of a particle in the magnetic field is smaller than the relativistic radiation beaming angle $\sim1/\gamma$, that is $\alpha\ll\Delta\theta$. In this case, 
the velocity ${\bf v}$ of a particle is almost constant whereas its acceleration ${\bf w\equiv\dot v}$ varies with time. Calculating the Fourier component of the electric field using the Li\'enard-Wiechart (retarded) potentials, one arrives at the following expression for the total energy 
emitted per unit solid angle $d\Omega$ per unit frequency $d\omega$:
\beq
dW=\frac{e^2}{2\pi c^3}\left(\frac{\omega}{\omega'}\right)^4
\left|{\bf n}\times\left[\left({\bf n}-\frac{\bf v}{c}\right)\times
{\bf w}_{\omega'}\right]\right|^2 d\Omega\,\frac{d\omega}{2\pi},
\label{dW-1}
\eeq
where ${\bf w}_{\omega'}=\int{\bf w}e^{i\omega't}\,dt$ is the Fourier component of the particle's acceleration, $\omega'=\omega\left(1-{\bf n\cdot v}/c\right)$, and ${\bf n}$ is the unit vector pointing towards the observer. We need to express the temporal Fourier component of the Lorentz acceleration, ${\bf w}=(e/\gamma m c){\bf v\times B}$, taken along the particle trajectory in terms of the Fourier component of the field in the spatial and temporal domains \citet{M06}. In the static case, i.e., when the magnetic field is independent of time, the ensemble-averages acceleration spectrum reads:
\begin{eqnarray}
\langle|{\bf w}_{\omega'}|^2\rangle
&=&(2\pi V)^{-1}\int|{\bf w}_{\bf k}|^2\delta(\omega'+{\bf k\cdot v})\,
d{\bf k},
\label{w1s}\\
|{\bf w}_{\bf k}|^2 
&=& (ev/\gamma m c)^2 (\delta_{\alpha \beta} - v^{-2} v_\alpha v_\beta)\,
V K_{\alpha \beta}({\bf k}),
\label{w2s}
\end{eqnarray}
$K_{\alpha \beta}({\bf r},t)
= T^{-1}V^{-1}\int B_\alpha({\bf r}',t') B_\beta({\bf r'+r},t'+t)
\,d{\bf r}' dt$ is the second-order correlation tensor of the magnetic field.

We adopt the following geometry: the Weibel current filaments are aligned with the $z$ direction and their magnetic fields lie predominantly in the $x-y$ plane. For a GRB, this geometry corresponds to a shock which is located in the $x$-$y$-plane and is propagating along $z$-direction. As the shock is propagating through a medium, the produced field is transported downstream (in the shock frame) whereas new field is continuously generated at the shock front. Thus, the field is also random in the parallel ($z$) direction. Similar structure of the magnetic fields is expected in the laboratory experiments with the field being random in the plane perpendicular to the beam propagation direction. Thus, Weibel turbulence shall be {\em highly anisotropic}. Both the theoretical considerations and realistic 3D simulations of relativistic shocks indicate that the dynamics of the Weibel magnetic fields in the shock plane and along the normal to it are decoupled. Hence, the Fourier spectra of the field in the $x-y$ plane and in $z$ direction are independent. Thus, for the Weibel fields at shocks, the correlation tensor has the form
\beq
K_{\alpha \beta}({\bf k})=C(\delta_{\alpha \beta}-s_\alpha s_\beta)
f_z(k_\|) f_{xy}(k_\perp),
\eeq
where ${\bf s}$ is the unit vector along the filaments (and normal to the shock front in the GRB case), $C$ is the normalization constant proportional to $\langle B^2\rangle$, $f_z$ and $f_{xy}$ are the magnetic field spectra along ${\bf s}$ and in the perpendicular plane, respectively, $k_\bot=(k_x^2+k_y^2)^{1/2}$ and $k_\|=k_z$, and finally, the tensor $(\delta_{\alpha \beta}-s_\alpha s_\beta)$ is symmetric and its product with ${\bf s}$ is zero, implying orthogonality of $\bf s$ and $\bf B$.

Numerical simulations \citep{Fred+04} also indicate that the field transverse spectrum, $f_{xy}$, is well described by a broken power-law with the break scale comparable to the skin depth. We expect that the 
spectrum $f_z$, has similar properties. Therefore, we use the following models:
\beq
f_z(k_\|)=\frac{k_\|^{2\alpha_1}}{(\kappa_\|^2+k_\|^2)^{\beta_1}}, \quad 
f_{xy}(k_\bot)
=\frac{k_\perp^{2\alpha_2}}{(\kappa_\perp^2+k_\perp^2)^{\beta_2}},
\label{f}
\eeq 
where $\kappa_\|$ and $\kappa_\perp$ are parameters determining the location of the peaks in the spectra, $\alpha_1,\ \alpha_2,\ \beta_1,\ \beta_2$ are power-law exponents below and above a spectral peak ($\beta_1>\alpha_2+1/2$ and $\beta_2>\alpha_2+1$, for convergence at high-$k$).
 
We now evaluate Eqs. (\ref{w1s}),(\ref{w2s}). The scalar product of the two tensors is
\beq
(\delta_{\alpha \beta} - v_\alpha v_\beta/v^2)
(\delta_{\alpha \beta}-s_\alpha s_\beta)=1+(s_\alpha v_\alpha)^2/v^{2}
=1+\cos^2\Theta,
\eeq
where we used that $\delta_{\alpha \alpha}=3$. Here $\Theta$ is the angle between the filament direction (and normal to the shock for GRBs) and the particle velocity (in an observer's frame), which is approximately the direction toward an observer, that is ${\bf v\|k}$ for an ultra-relativistic particle (because of relativistic beaming, the emitted radiation is localized within a narrow cone of angle $\sim 1/\gamma$). Eq. (\ref{w1s}) becomes
\beq
\langle|{\bf w}_{\omega'}|^2\rangle
=\frac{C}{2\pi}\,(1+\cos^2\Theta)\int\!\! f_z(k_\|) f_{xy}(k_\perp)
\delta(\omega'+{\bf k\cdot v})\,dk_\|d^2 k_\perp.
\label{w-main}
\eeq
Equations (\ref{dW-1}),(\ref{w-main}) fully determine the spectrum of jitter radiation from relativistic electrons propagating through the Weibel turbulence.

\subsection{Isotropic electron distribution: the GRB case}

First, we can simplify the vector expression in (\ref{dW-1}). Indeed, in the ultrarelativistic case, the longitudinal component of the acceleration is small compared to the transverse component, $w_\|/w_\bot\sim1/\gamma^2\ll1$. Therefore ${\bf v}$ and ${\bf w}$ are approximately perpendicular to each other. Second, the dominant contribution to the integral over $d\Omega$ comes from small angles 
$\theta\sim1/\gamma$ with respect to the particle's velocity. Therefore, we approximately write $\omega'\simeq\omega\left(1-v/c+\theta^2/2\right) \simeq\frac{1}{2}\omega\left(1-v^2/c^2+\theta^2\right)
=\frac{1}{2}\omega\left(\theta^2+\gamma^{-2}\right)$. We now can replace integration over the solid angle $d\Omega\simeq\theta\,d\theta\,d\phi$ with integration over $d\phi\,d\omega'/\omega$ and integrate equation (\ref{dW-1}) over the azimuthal angle, $\phi$, from $0$ to $2\pi$. The angle-averaged spectral power emitted by a relativistic particle moving through small-scale random magnetic fields, under the assumption that the deflection angle is negligible and the particle trajectory is a straight line, has been derived elsewhere \citep{LL,M00}:
\beq
\frac{dW}{d\omega}=\frac{e^2\omega}{2\pi c^3}\int_{\omega/2\gamma^2}^\infty
\frac{\left|{\bf w}_{\omega'}\right|^2}{\omega'^2}
\left(1-\frac{\omega}{\omega'\gamma^2}+\frac{\omega^2}{2\omega'^2\gamma^4}
\right)\,d\omega' .
\label{dW/dw}
\eeq
 
\begin{figure}
\epsfig{file=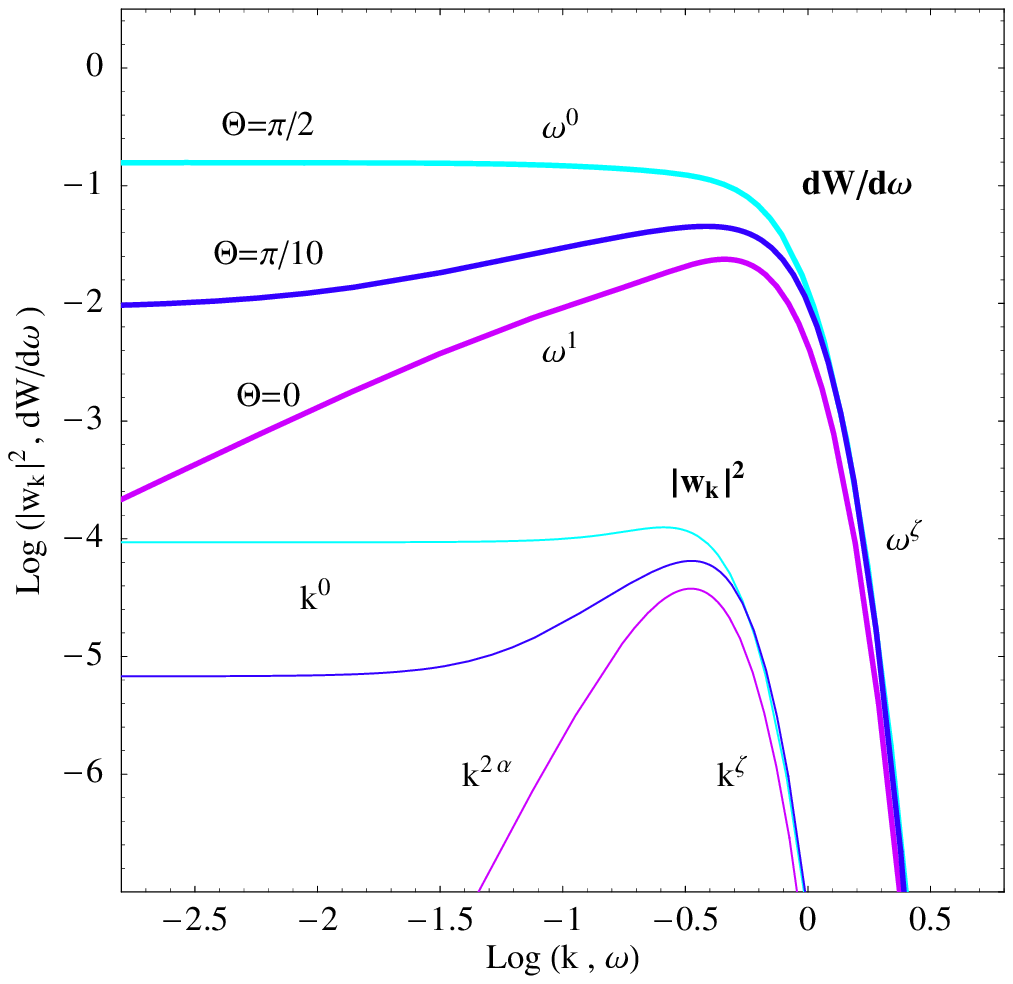,width=2.8in}
\caption{The $\log-\log$ plots of $|{\bf w_k}|^2$ vs $k$ ({\it thin lines}) 
and of $dW/d\omega$ vs $\omega$ ({\it thick lines}), for three viewing 
angles $\Theta=0,\ \pi/10,\ \pi/2$. The axes units are arbitrary. 
In this calculation we used $f_z=f_{xy}$ with 
$\alpha=2,\ \beta=20,\ \kappa=10,\ v=1$. The exponent 
$\zeta=\zeta(\alpha,\beta)$ is model dependent. We also chose $\gamma=1$ 
in order to align the peaks of $|{\bf w_k}|^2$ and $dW/d\omega$. 
Note that the actual peaks are at values $k,\ \omega$ lower than 
10 by a factor two or three. Note also that the spectrum $dW/d\omega$ 
levels off at oblique angles at frequencies much smaller than 
$\kappa v\gamma^2\sin\Theta$, whereas $|{\bf w_k}|^2$ indeed 
starts to flatten at $k\sim\kappa v\sin\Theta$. 
\label{f:2} }
\end{figure}

A spectrum from a shock viewed at an arbitrary angle, $0\leq\Theta\leq\pi/2$, is illustrated in Figure \ref{f:2}, which represents full numerical solutions of Eqs. (\ref{dW/dw}), (\ref{f}), (\ref{w-main}) for three different 
viewing angles. In calculation of $dW/d\omega$, the emitting electrons were assumed monoenergetic, for simplicity.  An important fact to note is that the jitter radiation spectrum varies with the viewing angle. When filaments are (and a shock velocity is, in the GRB case) along the line of sight, the low-energy spectrum is hard $F_\nu\propto\nu^1$, harder than the ``synchrotron line of death'' ($F_\nu\propto\nu^{1/3}$). As the viewing angle increases, the spectrum softens, and when the filaments are orthogonal to the line of sight, it becomes $F_\nu\propto\nu^0$. Another interesting feature is that at oblique angles, the spectrum does not soften simultaneously at all frequencies. Instead, there appears a smooth spectral break, which position depends on $\Theta$. The spectrum approaches $\sim\nu^0$ below the break and is harder above it.

\subsection{Beam electron distribution: the lab case}

Unlike a GRB shock, the distribution of radiating electrons is anisotropic in most of lab experiments. In particular, to diagnose the Weibel turbulence in the Hercules experiment \citep{GRB+Hercules08}, it has been suggested to launch a probe, nearly monoenergetic electron beam through plasma with the Weibel fields. Thus, the geometrical shape and the electron energy-momentum distribution of the probe beam are important. Here we assume the electrons to be monoenergetic and neglect the geometrical divergence of the beam, for simplicity (in the experiment, it is likely somewhat smaller than the relativistic beaming cone of $1/\gamma$ anyway). 

We again start from the Li\'enard-Wiechart potentials and the expression for the emitted power (\ref{dW-1}). We adopt the geometry such that the unit vector $\bf s$ is along the filaments, $\bf v$ is the particle's velocity and the unit vector $\bf n$ is toward an observed. We also define the unit vector $\hat{\bf v}={\bf v}/v$ and $\beta$. Unlike the isotropic case, we do not neglect terms proportional to $|\bf n\cdot w_{\omega'}|^2$ although they are small compared to those proportional to $\bf |w_{\omega'}|^2$. 
Using a similar approach, we arrive at the following expression for the emitted power per frequency, per solid angle, per electron:
\begin{eqnarray}
\frac{dW_\omega}{d\omega\,d\Omega}&=&\frac{e^2}{(2\pi)^2c^3}
\frac{1}{\left(1-\beta({\bf n\cdot\hat v})\right)^4}\nonumber\\
&\times&\left(\frac{e\beta}{\gamma m}\right)^2\beta^2\,\frac{C}{2\pi}
\int f_z(k_\|)f_{xy}(k_\bot)\delta\left(\omega\left(1-\beta({\bf n\cdot\hat v})\right)-{\bf k\cdot v}\right)\ dk_\|d^2k_\bot
\nonumber\\
&\times&\left\{ ({\bf n\cdot \hat v})^2\left(1+({\bf s\cdot\hat v})^2\right)+
\left[ ({\bf s\cdot n})^2+({\bf s\cdot\hat v})^2-2({\bf s\cdot n})({\bf s\cdot\hat v})({\bf n\cdot\hat v}) \right]
\right\}.
\end{eqnarray} 

One can readily see that the emitted power along the electron probe beam, i.e., when ${\bf n\cdot\hat v}=1$, is proportional to the electron acceleration spectrum:
\beq
\frac{dW_\omega}{d\omega\,d\Omega}\propto\bf |w_{\omega'}|^2,
\eeq
which is represented by the integral over $\bf k$, up to a constant. Thus, the electron beam can directly probe and diagnose the structure of the magnetic field distributions: $f_z$ if the beam is aligned with the Weibel filaments and $f_{xy}$ if the filaments are probed ``edge-on''; see Figure \ref{f:2} for the $\bf w$-spectrum. The angular shape of the radiation pattern (i.e., the angular distribution of intensity) can also be readily calculated by integrating over the frequency:
\beq
\frac{dW}{d\Omega}\propto
\left(1-\beta({\bf n\cdot\hat v})\right)^{-5}\propto \left(1+(\gamma\vartheta)^2\right)^{-5}.
\eeq 
The latter expression is valid for small angles $\vartheta$ between the probe beam direction and the line of sight.

\section{Conclusions}

The jitter spectra can deliver much information on the structure of the Weibel magnetic fields. As one can see, the spectrum depends on the spatial spectra of the magnetic fields modeled by Eqs. (\ref{f}). In particular, when viewing angles are 0 and 90 degrees, the contributions of the parallel and transverse magnetic field spectra are  decoupled. For instance, for $\theta=0$, the peak of the jitter radiation spectrum and its high-energy asymptotic slope are uniquely determined by the parallel correlation length 
$\kappa_\|$ and the large-$k$ magnetic field spectrum slope $k^\eta$ with $\eta=2\alpha_1-2\beta_1$. Similarly, the transverse jitter spectrum (at $\theta=\pi/2$) allows one to deduce these parameters for the transverse magnetic field spectrum, $f_{xy}$. At intermediate angles, one can determine the relative orientation of the current (and magnetic) filaments in the target and the radiation detector. 

It seems feasible to obtain jitter radiation in a laser-plasma experiment, such as Hercules. The Weibel turbulence to be studied will have much in common with the upstream region of a gamma-ray burst collisionless relativistic shock. It may be so even up to and at the main shock compression, where the Weibel filaments are present. In the downstream region the filaments are destroyed and the fields are significantly isotropized, as follows from simulations \citep{Spitk08}. In such a turbulence state, jitter radiation can still be present, but it will produce more isotropic, relatively soft spectra resembling those at $\theta\sim\pi/2$.

This work has been supported by NASA grants NNX07AJ50G, NNX08AL39G, NSF grant AST-0708213, and DOE grant  DE-FG02-07ER54940.

\clearpage

\end{document}